# Regulation and the Integrity of Spreadsheets in the Information Supply Chain

*Ralph Baxter,*
*ClusterSeven, 10 Fashion Street, London E1 6PX*
*rbaxter@clusterseven.com*

**ABSTRACT**

*Spreadsheets provide many of the key links between information systems, closing the gap between business needs and the capability of central systems. Recent regulations have brought these vulnerable parts of information supply chains into focus. The risk they present to the organisation depends on the role that they fulfil, with generic differences between their use as modeling tools and as operational applications.*

*Four sections of the Sarbanes-Oxley Act (SOX) are particularly relevant to the use of spreadsheets. Compliance with each of these sections is dependent on maintaining the integrity of those spreadsheets acting as operational applications. This can be achieved manually but at high cost. There are a range of commercially available off-the-shelf solutions that can reduce this cost. These may be divided into those that assist in the debugging of logic and more recently the arrival of solutions that monitor the change and user activity taking place in business- critical spreadsheets. ClusterSeven provides one of these monitoring solutions, highlighting areas of operational risk whilst also establishing a database of information to deliver new business intelligence.*

## 1. INTRODUCTION

Section 2 of this paper describes the role that spreadsheets play in business-critical operational environments, whilst Section 3 discusses the difficulty of eliminating them. This leads to an understanding of the generic risks presented by spreadsheets in Section 4 and the implications for the control requirements of SOX (Section 5). Section 6 then addresses the aspects of integrity that must be managed, with Section 7 describing the technical solutions that can assist.

## 2. SPREADSHEETS IN THE INFORMATION SUPPLY CHAIN

Recent regulation has focused attention on the way that information technology supports end-to-end business processes (for example, PricewaterhouseCoopers, 2004). (In this context an end-to-end business process is seen as a set of business activities that complete the loop from the first initiation of an activity (e.g. a customer placing an order) to the closure of that activity (e.g. cash in the bank).

This commonly demonstrates that the 'marriage' between systems and processes is not perfect (see Figure 1). Instead of one end-to-end IT system (or integrated set of systems) supporting the whole process, there are usually gaps where information is retrieved from one or more systems and manipulated before being transferred to the next system in the process.

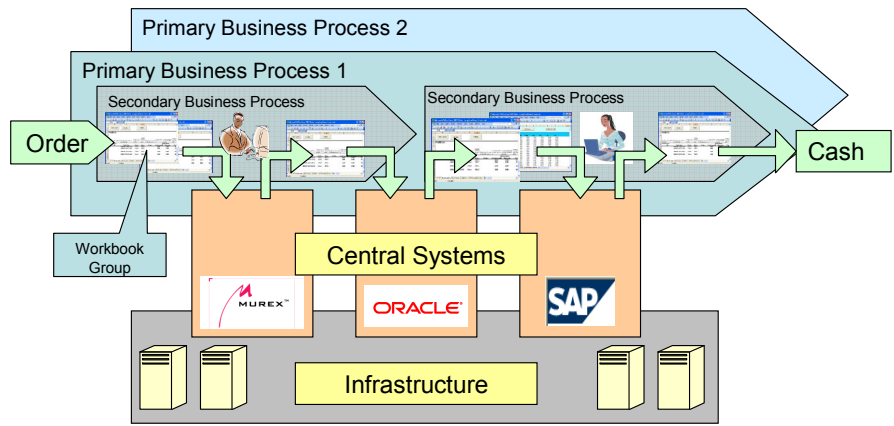

**Figure 1: The Relationship between Information Systems and Business Processes**

The management of information in the gaps between formal systems is clearly critical to the integrity output of the information chain – it being only as strong as its weakest link. End user computer applications, particularly spreadsheets tend to be one of the most pervasive support tools at these points (Howard, 2005). It is these spreadsheets – used as operational applications (i.e. supporting regular business tasks) – that are the main regulatory focus.

## 3. WHY DO SPREADSHEETS PERSIST?
Given the large technology investments made by businesses it begs the question as to why they continue to use spreadsheets to support their information supply chains. The reality is that central systems can only be as good as their original specifications. Since business needs are always changing, the lag between specification and roll out will be at least 6 months (and commonly much more). Hence, in practice, the central systems will always be out of date.

**Figure 2: The Evolution of Business Systems Functionality**

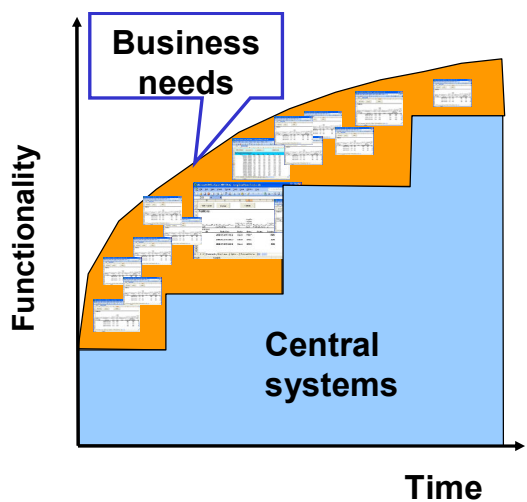

The difference between business needs and central systems capability (Figure 2) is most commonly filled by spreadsheets. The number of these applications reflects the rate of evolution of business needs (more change equals more spreadsheets) and the age of, and investment in, existing systems (older systems and less recent investment leads to more spreadsheets). Indeed, most new financial functionality is templated in spreadsheets before migration to formal systems – which may take years. It is therefore not surprising that much of the competitive edge of businesses (e.g. new financial products) is held in spreadsheet applications.

These aspects of business practice have been implicitly accepted for many years. However, the growth of regulatory concern has brought them to the attention of compliance and audit functions. This can result in policies to eliminate spreadsheets through migration to central systems. Inevitably, however, some are too volatile to be immediately or too complex to be replaced on a cost-effective basis. Other policies have attempted to ban the use of spreadsheets as part of business-critical activity – only to find that the business then does it anyway. The reality is that for many organisations elimination for any significant period of time is nearly impossible to achieve.

Other solutions must be found that can satisfy control requirements for risk management without eliminating the business benefits of spreadsheets. It is therefore necessary to understand the way in which spreadsheets are used in large organizations.

**4. SPREADSHEET RISK**
In understanding the business use of spreadsheets it is instructive to consider the difference between their use as a modeling tool versus their use as an operational business application. In reality these aspects represent end members of a spectrum but such a differentiation helps to identify the kind of risks that will exist and the most efficient way of mitigating them.

**Table 1: Comparison between Spreadsheet Usage as Modeling Tools versus Operational Applications**

|  | **Modeling Spreadsheets** | **Operational Spreadsheets** |
|---|---|---|
| **User** | Typically built and used by the same individual | Typically built by developers before being moved across to other individuals or parts of the business for usage. |
| **Persistence** | Models (often extremely complex) may be built over days or weeks, only to be made redundant soon after the relevant business decision has been made. | Models, both simple and complex become part of the critical information flow of the business. They commonly persist for many months or years. |
| **Structural/Functional Volatility** | High. There are likely to be substantial structural revisions from one day to the next as major elements of the model are added or replaced. | Low to medium. All of the key structural elements are likely to be in place. Further evolution of the business process will require maintenance changes – but only rare structural overhauls. |
| **Data Volatility** | Medium. Primarily related to the exploration of alternative scenarios. | High. As the application is relatively mature the transactional data becomes the key variables within the spreadsheet. |
| **Usage** | Likely to be intensive for a short period of time. Usage commonly restricted to a single individual or a small close-knit modeling team. | Usage will depend on the individual business process being supported. Some will be hourly or daily. Others may only feature at week or month end. Usage will involve handover between multiple individuals fulfilling different tasks – not necessarily in the same department. |

Table 1 illustrates that the primary risks in spreadsheets used for modeling are related to potential logic flaws in the creation of the workbook. However, in these business environments the user commonly has a good understanding of the spreadsheet structure and of what answers 'make sense'. Since decisions are based on multiple iterative scenarios the risks of decisions based on incorrect processing are relatively low. For these reasons modeling spreadsheets receive less attention from regulators (Buckner, 2004).

In contrast, the risks attached to operational spreadsheets depend on the *ongoing maintenance* of the logic integrity in the spreadsheet. These risks are increased by such factors as multiple users and their lack of detailed knowledge of the spreadsheet structure. Risks are also increased where the output is the aggregation of many transactions where it is unlikely that anybody has a good understanding of what the 'right number' ought to be, even though it may be a key input to financial control processes. It is for these reasons that operational spreadsheets are much more the focus of regulatory concern.

## 5. REGULATORY REQUIREMENTS

Sarbanes-Oxley represents just one face of the most recent focus on the operational risk and financial reporting of corporations. Whilst the specific requirements of each piece of regulation may be open to interpretation, the overall direction is clear: to ensure that businesses understand what is happening in their organisation; to be able to respond in a timely fashion to issues when they emerge; to have procedures in place that minimize the possibility of things going wrong in the first place; and, most personally, to hold the business and its key executives to account if they don't do it.

Executives will have to gain a deeper understanding of internal controls because their business decisions will be placing greater reliance on adequate internal controls and ensuring that they are deployed, maintained, adjusted and reported on as required. John Flaherty, former COSO chairman, said that this means "… that every division in a company needs to have a documented set of internal rules that control how data is generated, manipulated, recorded and reported …"

For SOX there are four sections most relevant to spreadsheets and their controls:

**Table 2: SOX Implications for Operational Spreadsheets**

| Section | Requirements | Spreadsheet Implications |
|---|---|---|
| **103** **Auditing, quality control, and independence standards** | Independent auditors must include an evaluation of the Company's internal controls in their report. The evaluation will include a description of material weaknesses in internal controls and material non-compliance with them. | Un-monitored spreadsheets in the critical information supply chains will fail this test. |
| **302** **Corporate responsibility for financial reports** | Executives must evaluate the effectiveness of internal controls every quarter. Financial reports must include their conclusions about internal controls and explain any significant changes to them. All frauds, no matter how small, must be disclosed to the Company's auditors and to the Audit Committee of the Company's Board of Directors. | It may be possible to eliminate spreadsheets temporarily but unlikely they can be eliminated for every quarterly report. Spreadsheets are also a common source of fraud. |

| 304<br>**Forfeiture of certain bonuses and profits** | Accounting restatements due to material noncompliance of the Company with reporting requirements of securities laws, and that are the result of misconduct, could result in the Executives having to reimburse the Company for their bonuses, or for any profits they realize from the sale of Company securities. | Spreadsheet errors have been the source of material financial mis-filings that would result in triggering this clause. |
|---|---|---|
| 404<br>**Management assessment of internal controls** | The SEC requires that the annual report contain an internal control report, which:<br>• States management's responsibility for establishing and maintaining an adequate internal control structure and procedures for financial reporting; and,<br>• Assesses, as of the end of the most recent fiscal year of the Company, the effectiveness of the internal control structure and procedures for financial reporting.<br>The Company's auditor is required to attest to and report on management's assessment of internal controls. | Un-monitored spreadsheets in the critical information supply chains will fail this test and are likely to result in qualified statements of control. |

It is clear from the risks identified above that spreadsheets have potentially significant failings against regulatory demands and more general tests of business control:

- They are highly vulnerable to error and, occasionally, fraud
- The information they contain and the user interaction with them are not transparent to the rest of the organisation.
- It takes significant time and effort to understand unexpected changes and to respond and communicate them as appropriate

In order to resolve these challenges organizations must use processes and technology that can ensure the integrity of business critical spreadsheets . Solutions may vary from entirely manual to strongly technologically enabled but all must focus on the possible causes of losing integrity of the spreadsheet output.

**6. WHAT GOVERNS INTEGRITY?**

Full spreadsheet integrity (i.e. assurance that the output is the expected processing of the input) is dependent on five key elements:

1. That the programmer of the spreadsheet logic model correctly understands the transactional process to be implemented (i.e. correct specification)
2. That the programmer has created the required logic without errors (i.e. no bugs)
3. That subsequent data inputs are valid (whether manual or automatic)
4. That subsequent user and maintenance activity does not corrupt the original logic.
5. That where multiple user tasks are performed on a spreadsheet these are performed in the correct order.

Although these challenges are relatively short to define the pervasiveness of spreadsheets and their almost infinite flexibility means that solutions have taken much longer to emerge.

It is also apparent that the 'culture' around spreadsheet within many large organizations does not contribute to avoiding or removing these problems. For example spreadsheets are commonly seen as a temporary solution that will be replaced at the appropriate time by 'proper' investment in fully architected systems – and by implication it is not sensible to make this investment in the spreadsheet application itself. This applies even in organizations where some spreadsheet models have persisted for many years. It is ironic that spreadsheets are often viewed as a tactical solution when they are one of the longest standing parts of most enterprise information systems.

## 7. HOW CAN TECHNOLOGY HELP?

Several tools have been created to address the potential flaws in logic creation. These include Spreadsheet Professional, HMC&E SpACE, Operis OAK, recent Microsoft Excel 2003 error tips and others. Clearly the wide variety of spreadsheet structures and logic requirements means that none of these tools can be 100% effective. They are reliant on looking for inconsistencies in successive cell formulae or performing checks for logic elements that are known to be particularly error prone (e.g. nested IF formulae).

Despite the research demonstrating the prevalence of errors in almost all spreadsheet logic it is clear that with (and without) logic-checking tools most organizations feel comfortable that they can get their operational spreadsheets to be reliable at particular points of time (e.g. testing and audit). The next question is how to maintain this quality after the spreadsheet enters operational usage.

Perhaps the most common solution is to impose some form of lockdown on spreadsheet change. This can be effective in highly resourced environments (where developers are on hand to change, test and re-issue a revised version within the timetable of business needs) or where the spreadsheet application has become very mature in its usage (i.e. business needs are not changing). However, in less resource-rich or mature environments this policy inevitably fails because it prevents the user exercising their own business knowledge to resolve their own problems. As a result it is almost always circumvented.

A second option is to continue using the logic tools in the operational environment. However, these tools are usually inappropriate for such use as they are designed to look for logic inconsistencies rather than track broader categories of data behaviour. Moreover, they would have to be utilized after every user interaction (probably impractical) and must be interpreted by someone who understands the underlying structure of the spreadsheet (less common in operational spreadsheet usage).

A third option is now appearing. These are solutions designed to specifically address operational usage of spreadsheets. ClusterSeven is one example of these new solutions that focuses on the change management and user interaction with business critical spreadsheets. In so doing ClusterSeven can expose all of the time variant information contained within spreadsheets – be it data, functionality or usage. This allows it to highlight areas of operational risk (when activity does not conform to expected patterns) and also to create business intelligence (such as reporting on the trends of data values for particular key performance indicators).

Given the low take up of existing technology tools (compared to the pervasiveness of Excel) it is appropriate to ask whether yet more technology will provide the answer. ClusterSeven believes that a number of factors are now converging to make this a reality: firstly auditors and regulators are becoming increasingly vocal that the status quo is not satisfactory; secondly the combination of logic tools and operational tools enables the whole spreadsheet lifecycle to be managed and thirdly there is growing acceptance (all be it grudging in places) that spreadsheets are here to stay, necessitating a strategic approach to the problem.

## 8. CONCLUSIONS

- Spreadsheets populate the information supply chains of many large organizations.
- As the 'weakest link' in the information supply chain they have become the target of regulatory concerns about financial reporting.
- Regulatory concerns can be addressed by adopting processes that ensure the continuing integrity of key operational spreadsheets.
- Integrity can be maintained through the application of arduous manual processes or via the assistance of technology.
- Integrity also requires a shift in the culture of organizations to see spreadsheet technology as a persistent strategic part of their infrastructure rather than a short term tactical fix.